\documentclass[%preprint,
nofootinbib,
prd,
amsmath,amssymb,
aps,
floatfix,
]{revtex4-2}

\usepackage{graphicx, color}% Include figure files
\usepackage{dcolumn}% Align table columns on decimal point
\usepackage{bm}% bold math
\usepackage[mathlines]{lineno}% Enable numbering of text and display math
\usepackage{amsmath}
\usepackage{amssymb}
\usepackage{float}
\usepackage{caption}
\usepackage{mathrsfs}
\usepackage{bbold}
\usepackage{lipsum}
\usepackage{hyperref}
\usepackage{slashed}
\usepackage{soul}
\usepackage{tcolorbox}
\usepackage[normalem]{ulem}
\hypersetup{
    colorlinks=true,       % false: boxed links; true: colored links
    linkcolor=red,          % color of internal links
    citecolor=blue,        % color of links to bibliography
    filecolor=magenta,      % color of file links
    urlcolor=blue           % color of external links
}
\usepackage[all]{hypcap} %Link navagates to top of figure instead of caption (below fig)
\usepackage{cleveref}
\usepackage[]{physics}
\usepackage{hyperref}% add hypertext capabilities
\usepackage[mathlines]{lineno}% Enable numbering of text and display math
\usepackage{makeidx}

\definecolor{customgray}{RGB}{230,230,230} % Define a custom light gray

\newcommand{\bea}{\begin{eqnarray}}
\newcommand{\eea}{\end{eqnarray}}

\newcommand{\x}{\mathbf{x}}
\newcommand{\vel}{\mathbf{v}}
\newcommand{\xphys}{\mathbf{r}}
\newcommand{\vflow}{\mathbf{v}_{\rm flow}}
\newcommand{\vpec}{\mathbf{v}_{\rm pec}}

\newcommand{\vsim}{\mathbf{p}}
\newcommand{\phiN}{\phi_{\rm N}}
\newcommand{\rhobar}{\bar{\rho}}

\newcommand{\fnet}{\mathbf{f_{\rm net}}}
\newcommand{\fbg}{\mathbf{f_{\rm b.g.}}}
\newcommand{\fpert}{\mathbf{f_{\rm pert}}}
\newcommand{\fpertcom}{\mathbf{f_{\rm pert,com}}}
\newcommand{\GN}{G_{\rm N}}

\begin{document}

\title{Clustering of cosmic string loops within a Milky-way like halo}

\author{Itamar Allali}\email{iallali@nd.edu}
\affiliation{Department of Physics and Astronomy, University of Notre Dame, Notre Dame, IN 46556}
\affiliation{Department of Physics, Brown University, Providence, RI 02912, USA}
\author{Mudit Jain}\email{mudit.jain@rice.edu}
\affiliation{Theoretical Particle Physics and Cosmology, King’s College London, Strand, London, WC2R 2LS, United Kingdom}
\affiliation{Department of Physics and Astronomy, Rice University, Houston, Texas 77005, U.S.A.}

\author{Shi Yan}\email{shi\_yan@brown.edu}
\affiliation{Department of Physics, University of Wisconsin, Madison, WI 53707, USA}
\affiliation{Department of Physics, Brown University, Providence, RI 02912, USA}

\date{\today}% It is always \today, today,

\begin{abstract}
Loops of cosmic string experience a recoil from anisotropic gravitational radiation, known as the \textit{rocket effect}, which influences the extent to which they are captured by galaxies during structure formation. Analytical studies have reached different conclusions regarding loop capture in galaxies: early treatments argued for efficient capture, while later analyses incorporating the loop rocket force throughout halo formation found that capture efficiency is reduced and strongly dependent on loop size. In this work, we employ the N-body simulation code {\tt GADGET-4}, introducing non-backreacting tracer particles subject to a constant recoil force to model cosmic string loops with the rocket effect. We simulate the formation of a Milky-Way–like halo from redshift $z=127$ to $z=0$, considering loop populations characterized by a range of length parameters $\xi$, inversely proportional to the rocket acceleration. We find that the number of captured loops exhibits a pronounced peak at $\xi_{\rm peak}\simeq 12.5$, arising from the competition between rocket-driven ejection at small $\xi$ and the declining intrinsic loop abundance at large $\xi$. For fiducial string tensions, this corresponds to $\mathcal{O}(10^6)$ loops within the halo. We further find that loops with weak rocket forces closely trace the dark-matter distribution, while those subject to stronger recoil but still captured -- particularly the most abundant loops near $\xi_{\rm peak}$ -- are preferentially concentrated toward the central regions of the halo.

\end{abstract}

\maketitle
%\flushbottom
\tableofcontents

% ----------------- %
% ----------------- %
\section{Introduction}
\label{sec:intro}
% ----------------- %
% ----------------- %

Phase transitions throughout cosmic history can leave behind a variety of topological defects. In many physical theories, linear topological defects called cosmic strings are predicted. These strings evolve dynamically after formation, oscillating to produce gravitational radiation and forming loops of string when strings collide and reconnect. Cosmic strings and string loops, though not yet directly observed, can have a variety of potential observational signatures, including gravitational lensing, microlensing, and gravitational-wave emission 
\cite{Siemens:2006yp,Chernoff:2018kfv,DePies:2007sr,Khakhaleva-Li:2020wbr}. Also see \cite{Vilenkin:2000jqa,Vachaspati:2015cma,Chernoff:2014cba} for reviews.

For accurate detection prospects, it is important to understand the spatial distribution of oscillating loops of cosmic strings. Historically, it has been assumed that loops are distributed uniformly in space. The loops are formed at relativistic speeds, and although they are slowed by cosmic expansion, they experience a recoil force from their gravitational radiation. This anisotropic radiation is known as the ``rocket effect," giving the loops a large late-time acceleration, which has been argued to suppress or prevent their capture by gravitational structure formation. 

Some challenges to this paradigm have arisen, with some disagreement. If the loops are instead captured by galaxies, then one would expect a much higher density of loops near the Earth and drastically different observational predictions. This question was first addressed in \cite{Chernoff:2009tp}, suggesting that by the time the rocket effect is strong, the loops have slowed to essentially comoving speeds and subsequently fallen into dark matter (DM) haloes, at which point the rocket effect is not strong enough to eject them. Thus, it was argued, string loops are in fact captured in galaxies. In contrast, in a paper by some of the authors of the present work \cite{Jain:2020dct}, it was instead argued that fewer loops are captured when the rocket effect can prevent the infall of loops throughout the process of forming a galaxy. Both groups approached this issue with analytical methods, using a spherical top hat model of halo formation. The discrepancy between the two amounts to orders of magnitude, highlighting the sensitivity of analytic estimates to idealized assumptions about halo formation.

In this work, we aim to apply numerical methods to the questions of the number and the spatial distribution of captured loops of cosmic string today. We perform a set of DM-only N-body simulations of the formation of a Milky-Way-like galaxy, inserting particles experiencing a constant force to simulate the behavior of cosmic string loops with the rocket effect. The simulations result in a DM distribution of the Navarro-Frenk-White (NFW) type, a more accurate recreation of our galaxy compared to the spherical top-hat model. We analyze the results of the simulation to determine the expected capture and spatial distribution of string loops in Milky-Way–like galaxies. In these simulations, the string loops are treated as effectively massless tracer particles that do not back-react on the formation of the dark-matter halo, allowing us to isolate the impact of the rocket effect on their dynamics within a realistic galactic potential.

In \cref{sec:particl_EOM}, we construct the necessary equations of motion for the string loop evolution. First we construct evolution equations for the string loops, then review the rocket effect, and then recast the equations of motion appropriately for the simulations. \Cref{sec:sim_methods} describes the N-body simulation we use and the set of simulations that are performed for our analysis. In \cref{sec:results} we present the results of the simulations, and in \cref{sec:summary} we summarize.

%%%%%%%%%%
\section{Particle equations of motion}
\label{sec:particl_EOM}
%%%%%%%%%%

We will construct the equations of motion for string loops that experience the rocket effect in an expanding universe. To do so, we first delineate the equations governing the evolution of non-relativistic particles in a cosmological background, and then specialize to loops whose recoil from anisotropic gravitational-wave emission produces an additional acceleration.

\subsection{Evolution of particles}
\label{subsec:evol_particles}

Consider a universe with evolution dictated by a pressure-less matter fluid (cold DM, with equation of state parameter $w=0$) and a cosmological constant dark energy fluid. This universe is characterized via a standard Friedmann-Lemaitre-Robertson-Walker (FLRW) metric (under the assumptions of large-scale homogeneity and isotropy), which in comoving coordinates is characterized by the following differential metric
\begin{align} 
ds^2 = dt^2 - a^2(t)\,d\x^2. 
\end{align}
Here, $t$ denotes cosmic time, $a(t)$ the scale factor tracking the expansion of space, and $\x$ a three-vector denoting comoving coordinates ($x^i$ denoting one component), such that physical distances are given by $a(t) x^i \equiv r^i$. 

The equation of motion of any non-relativistic point particle in this background is
\begin{align}\label{e:Newtons_law}
    \frac{d{\vel}}{dt} = \fnet,
\end{align}
where $\vel$ is the physical velocity $\vel \equiv d\xphys/dt$ and $\fnet$ is the net force on the point particle per unit mass. This physical velocity, expressed in comoving coordinates, can be split into two pieces:
\begin{align} 
\vel =  d\xphys/dt = \dot{a}\x + a \dot{\x} \equiv \vflow + \vpec\,.
\end{align}
We refer to $\vflow\equiv \dot{a} \x$ as the ``Hubble flow" velocity (the velocity component that comoves with the expansion of the universe), and a ``peculiar" velocity $\vpec\equiv a\dot{\x}$ (the component that moves with respect to the homogeneous and isotropic background flow, changing the comoving coordinate $\x$). We use overdots to denote derivatives with respect to cosmic time $t$.

A test particle which co-moves with the Hubble flow has $\dot{\x}=0$ and experiences the usual Newtonian gravitational force $\fbg$ due to the homogeneous background matter density $\rhobar$. That is, $\fbg = -\nabla \phiN$ where 
\begin{align} 
-\nabla \phiN = -\GN \int_{V} {\rm d}^3{\xphys'}\,\rhobar \frac{\xphys-\xphys'}{|\xphys-\xphys'|^3} = -\frac{4\pi}{3}\GN \rhobar\,\xphys\,.
\end{align}
The particle's equation of motion $\ddot{\xphys} = \ddot{a}\x = \fbg$ precisely gives the Friedmann (acceleration) equation, governing the evolution of the scale factor in a matter dominated Universe
\begin{align} \frac{\ddot{a}}{a} =  -\frac{4\pi}{3}\GN \rhobar\,.
\end{align}

To study the evolution of a test particle which is not comoving, one may break up the total force $\fnet$ into two pieces $\fbg$ and $\fpert$, where $\fbg$ encodes the influence of the background ($-\nabla\phiN$ above), and $\fpert$ gives all additional forces which lead to perturbations around the background flow motion. From $\dot{\vel}=\fbg+\fpert=\ddot{a}\x + (2\dot{a}\dot{\x}+a\ddot{\x})$, we therefore have
\begin{align}
\label{eq:pertEOM1}
\ddot{\x}+2H\dot{\x} =\frac{\fpert}{a}\,,
\end{align}
where we have used the Hubble rate $H(t) = \dot{a}/a$. This, along with the anticipation of our application to the {\tt GADGET-4} simulation, lets us define a convenient velocity variable 
\begin{align} \vsim\equiv a^2 \dot{\x}.
\end{align} 
This is useful due to its time derivative $\dot{\vsim}=a^2 \ddot{\x}+2\dot{a}a\dot{\x}=a^2 (\ddot{\x}+2H\dot{\x})$, giving the simple equation of motion
\begin{equation}
    \dot{\vsim} = a \, \fpert \, .
\end{equation}
For the purposes of simulations, it is further useful to define a comoving force $\fpertcom = a^2\fpert$. Then, in terms of logarithmic time, $\tau\equiv \ln a$, we arrive at the following relevant phase space equations of motion for particles in the simulation box
\begin{align}
\label{eq:xandptauEOM}
\frac{d\x}{d\tau} &= \frac{\vsim}{a^2(\tau) H(\tau)}\nonumber\\
\frac{d\vsim}{d\tau} &= \frac{1}{a(\tau)H(\tau)}\fpertcom\,.
\end{align}
These provide the cosmological framework into which we will introduce the rocket acceleration.

\subsection{The Rocket effect}
\label{subsec:rocket}

Oscillating cosmic string loops emit gravitational waves anisotropically, producing a net recoil (the \emph{rocket effect}) that accelerates the loop center of mass \cite{Vachaspati:1984gt,Blanco-Pillado:2011egf}. We follow the notation of Ref.~\cite{Jain:2020dct} and characterize loop size by
\begin{align}
\label{eq:xi_define}
  \xi \;\equiv\; \frac{l}{l_\ast}\,, 
  \qquad l_\ast \;\equiv\; \Gamma\,G\mu\,t_0\,,
\end{align}
where \(l\) is the invariant loop length, \(\mu\) is the string tension, \(G\) is Newton's constant, \(t_0\) is the present cosmic time, and \(\Gamma \simeq 50\) is the dimensionless power coefficient for GW emission by loops. Loops of size \(l_{\ast}\), corresponding to \(\xi \rightarrow 1\), are those that survive until today.\footnote{Taking the gravitational wave radiation rate as \(\dot{E} \simeq \Gamma G\mu^2\), the lifetime of a loop of length \(l\) is \(\tau \simeq l/\Gamma G\mu\).} Since an overwhelming majority of captured loops are much larger and hence large lifetimes compared to $t_0$, we can safely neglect loop ejections (see~\cite{Jain:2020dct} and also see~\cref{fig:capture} ahead).
 
The recoil (thrust) due to anisotropic GW emission is \(F \sim \Gamma_p\,G\mu^2\), with \(\Gamma_p\) an \(\mathcal{O}(1\!-\!10)\) anisotropy parameter (empirically \(\Gamma_p/\Gamma \sim 0.1\) for typical loops, see \cite{Blanco-Pillado:2011egf}). Treating each loop as a point particle of mass \(m \sim \mu l\) and assigning it a fixed (time-independent) rocket direction \(\hat{\mathbf n}\) at creation, the resulting acceleration (or force per unit mass) magnitude is
\begin{align}
  f_{\rm rocket} \;\equiv\; \frac{F}{m} \;\simeq\; \frac{\Gamma_p\,G\mu}{l}
  \;=\; \frac{\Gamma_p}{\Gamma}\,\frac{1}{\xi\,t_0}\,,
  \label{eq:ar}
\end{align}
which is constant for a given \(\xi\).\footnote{In our simulations, loops are treated as point particles with prescribed rocket acceleration and \emph{no} shrinking of \(l\); \cref{eq:ar} is therefore implemented with fixed \(\xi\).}

The average differential number density of large loops (\(\xi \gg 1\)) in matter domination (\(3000 \gtrsim z \gtrsim 1\)), as dictated by Nambu-Goto simulations~\cite{Blanco-Pillado:2017oxo}, is
\begin{align}
\label{eq:number_density_actual}
    n(\xi,z) :=\frac{d\tilde{n}}{d\ln\xi} \simeq \frac{0.5 H_0^{3/2}\Omega_{r0}^{3/4}(1+z)^3}{l^{3/2}_{\ast}}\frac{1}{\xi^{3/2}}\, 
    \approx 10^{-6} (G\mu)^{-3/2}\xi^{-3/2} t_0^{-3} (1+z)^3\,,
\end{align}
where we have taken \(\Omega_{r0} = 9\times 10^{-5}\) for the present time radiation fraction. Loops are chopped off from the string network at mildly relativistic velocities, after which their peculiar velocity \(\mathbf v_{\rm pec}\) evolves as
\begin{align}
  \dot{\mathbf v}_{\rm pec} + H\,\mathbf v_{\rm pec} 
  \;=\; a_{\rm rocket}\,\hat{\mathbf n}
  \;=\; \frac{\Gamma_p}{\Gamma}\,\frac{1}{\xi\,t_0}\,\hat{\mathbf n}\,.
  \label{eq:vpec-eom}
\end{align}
In matter domination, \(H=2/(3t)\), \cref{eq:vpec-eom} integrates to
\begin{align}
  \mathbf v_{\rm pec}(t) 
  \;=\; \mathbf v_0(t) 
        + \frac{3}{5}\,\frac{\Gamma_p}{\Gamma}\,\frac{t}{\xi\,t_0}\,\hat{\mathbf n}\,,
  \label{eq:vpec-sol}
\end{align}
with the Hubble-damped contribution from the loop’s initial motion
\begin{align}
\label{eq:v0_damped_damped}
  \mathbf v_0(t) \;\simeq\; \mathbf v_f\,
  \Big(\frac{t_f}{t_{\rm eq}}\Big)^{\!1/2}
  \Big(\frac{t_{\rm eq}}{t}\Big)^{\!2/3}
  \;=\; 2.6\,(1+z)\,(\xi G\mu)^{1/2}\,\mathbf v_f\,,
\end{align}
where \(v_f\sim 0.3\) is the typical formation speed, \(t_f \sim 10\,\xi\,\Gamma G\mu\,t_0\) is the formation time (in the radiation era) of a loop of size \(\xi\), and \(t_{\rm eq}\) is the time of matter–radiation equality \cite{Jain:2020dct}.

The time (in matter dominated era) at which the rocket term overtakes the Hubble-damped term in \cref{eq:vpec-sol}, is found by equating the two contributions, giving
\begin{align}
  t_r \;\sim\; 1.2\left(\frac{\Gamma_p}{\Gamma}\right)^{-3/5}\xi^{9/10}\,(G\mu)^{3/10}\,t_0\,,
  \qquad
  1+z_r \;\sim\; 0.9\left(\frac{\Gamma_p}{\Gamma}\right)^{2/5}\,\xi^{-3/5}\,(G\mu)^{-1/5}\,.
  \label{eq:tr-zr}
\end{align}
For \(\xi \ll \xi_r\) (defined below) and small \(G\mu\), this transition typically occurs well \emph{before} halo virialization, and the velocity becomes rocket-dominated:
\begin{align}
  \mathbf v_{\rm pec}(t) \;\approx\; 
  \Big(\frac{3}{5}\,\frac{\Gamma_p}{\Gamma}\Big)\,\frac{t}{\xi\,t_0}\,\hat{\mathbf n}
  \;\approx\; 0.06\,\frac{t}{\xi\,t_0}\,\hat{\mathbf n}\,,
  \label{eq:vpec-rocket-dominated}
\end{align}
where the numerical coefficient uses \(\Gamma_p/\Gamma \approx 0.1\) as in \cite{Jain:2020dct}. 

A convenient way to express the onset of the rocket-dominated regime relative to the redshift of halo virialization \(z_v\) is via the critical size \(\xi_r\) obtained from \(z_r \gg z_v\):
\begin{align}
  \xi \ll \xi_r 
  \;\sim\; 2000\,\mu_{-15}^{-1/3}
  \Big(\frac{1+z_v}{4}\Big)^{-5/3}\,,
  \qquad \mu_{-15}\equiv \frac{G\mu}{10^{-15}}\,.
  \label{eq:xir}
\end{align}
In this regime, \cref{eq:vpec-rocket-dominated} is an accurate approximation at and before halo turnaround.

From \cref{eq:xi_define}, the size of loops are $l \sim \xi\times 2\times 10^{-10}\,\mu_{-15}$ Mpc, which is enormously small as compared to length scales appropriate for large scale structure. Furthermore, the velocities of these loops are highly non-relativistic, as seen from \cref{eq:v0_damped_damped}, $v_0 \sim \sqrt{\xi}\times 2.5\times 10^{-8}(1+z)\mu_{-15}^{1/2}$. Therefore, for the purposes of simulating string loop capture during large scale structure formation (where we will be only interested in their dynamics for redshifts smaller than $\mathcal{O}(130)$), it will be sufficient to treat them as point like particles that are subject to an additional ``rocket force" besides their dark-matter-particle-like behavior. 

\subsection{Implementation in simulations}

Within the particle-evolution framework derived in \cref{subsec:evol_particles}, the rocket acceleration simply appears as an additional force, supplementing the gravitational force from small inhomogeneities in neighboring particles, $\fpert$. In comoving form therefore, we make the replacement
\begin{align}
\fpertcom \rightarrow \tilde{\bf f}_{\rm com, pert} = \fpertcom + a^2 \mathbf{f}_{\rm rocket} = \fpertcom + a^2\left(\frac{\Gamma_p}{\Gamma}\frac{1}{\xi t_0}\right)\hat{\mathbf n}\,.
\end{align}
The steps for updating positions and velocities at a time $\tau$ to a time $\tau + \delta \tau$ are
\begin{align} 
\x(\tau+\delta \tau) &= \x(\tau) + \vsim\int_\tau^{\tau+\delta\tau}\frac{d\tau'}{a^2(\tau') H(\tau')}\nonumber\\
\vsim(\tau+\delta \tau) &= \vsim(\tau) + \tilde{\bf f}_{\rm com, pert} \int_\tau^{\tau+\delta\tau}\frac{d\tau'}{a(\tau') H(\tau')}\,.
\end{align}

%%%%%%%%%%
\section{Simulation methods}
\label{sec:sim_methods}
%%%%%%%%%%

We simulate the formation of a Milky Way-like galaxy utilizing the {\tt GADGET-4}\footnote{\label{gadget}https://wwwmpa.mpa-garching.mpg.de/gadget4/} N-body and hydrodynamical structure formation code \cite{Springel:2020plp}. We have modified the code minimally to include additional particles with a constant unidirectional force, in addition to gravitational interactions, to simulate cosmic string loops with the rocket effect. To be consistent with the usual $\Lambda$CDM cosmology, the masses of these new particles are chosen to be sufficiently small such that they do not back-react onto the structure formation. Our simulations also include dust grains with only gravitational interactions, simulating the formation of dark matter halos.

\begin{figure}
    \centering

    % --- top plot ---
    \includegraphics[width=0.5\linewidth]{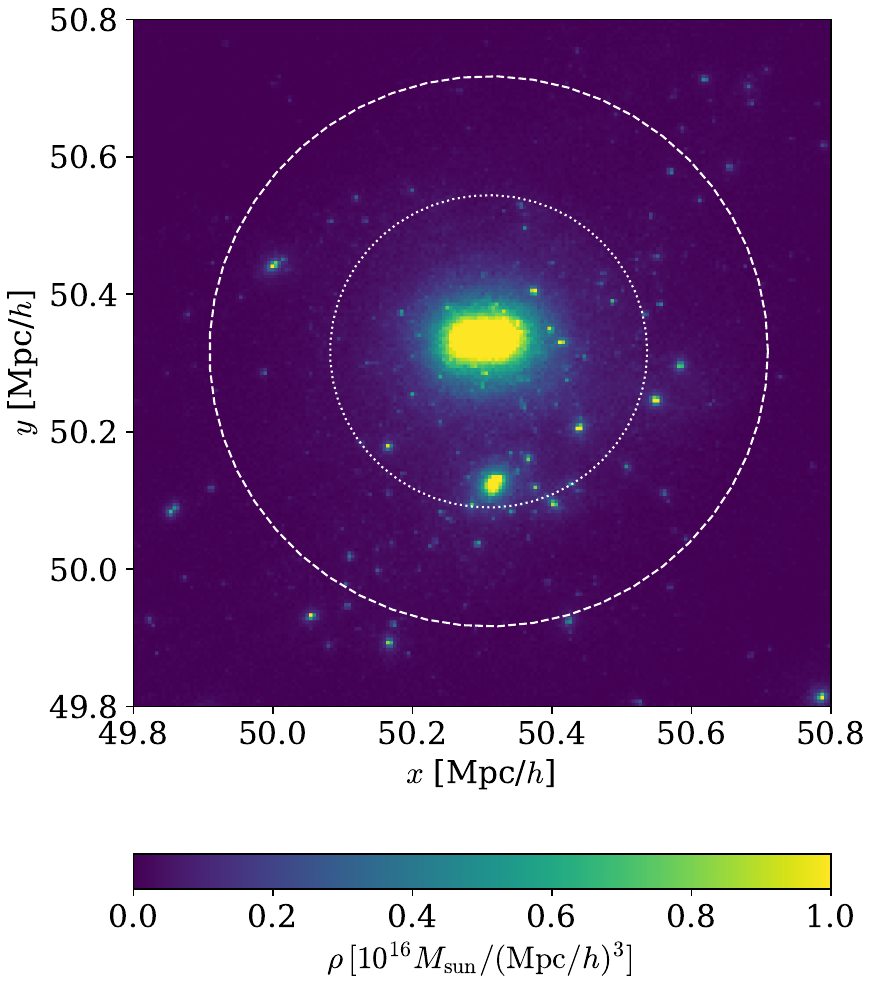}
    \caption{The dark matter density in two dimensions (restricting to a $2 \mbox{ Mpc}$ range for the third dimension) is shown for an example output of the N-body simulation. Mass density is indicated by color, given in $10^{16} M_{\text{sun}} / (\mbox{Mpc}/h)^3$. The white dashed (dotted) circle indicates the region within $r_{50}\approx 0.4 \mbox{ Mpc}/h$ ($r_{200}\approx0.227 \mbox{ Mpc}/h$) from the center of mass of the main halo.}
    \label{fig:sim}

    \vspace{0.7em} % small vertical gap

    % --- bottom plot: two panels side by side ---
    \includegraphics[width=0.48\linewidth]{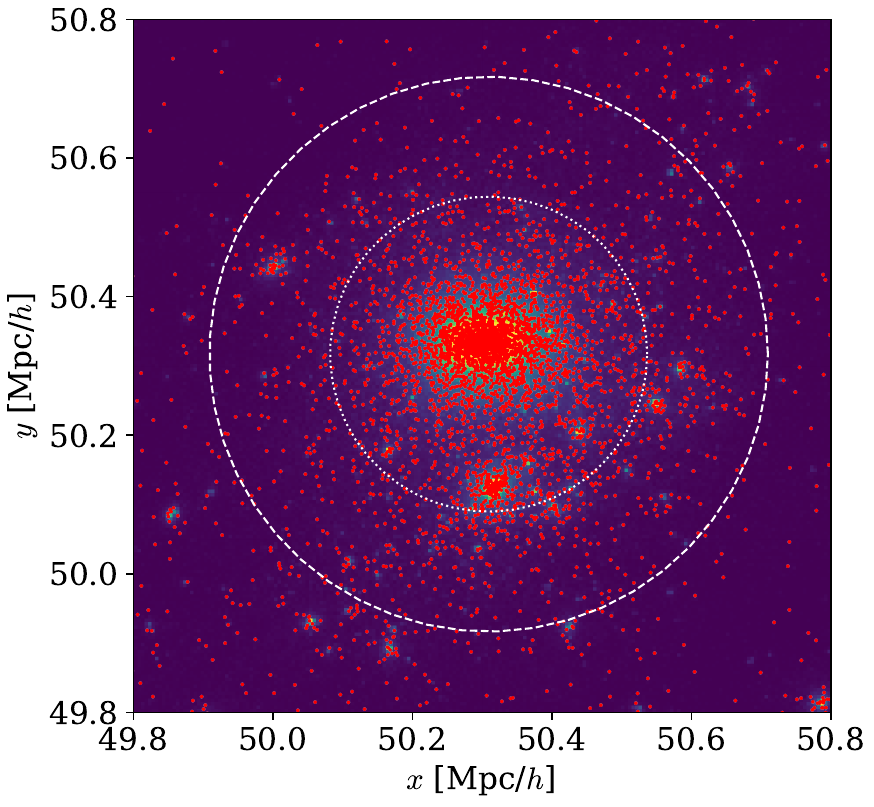}
    \includegraphics[width=0.48\linewidth]{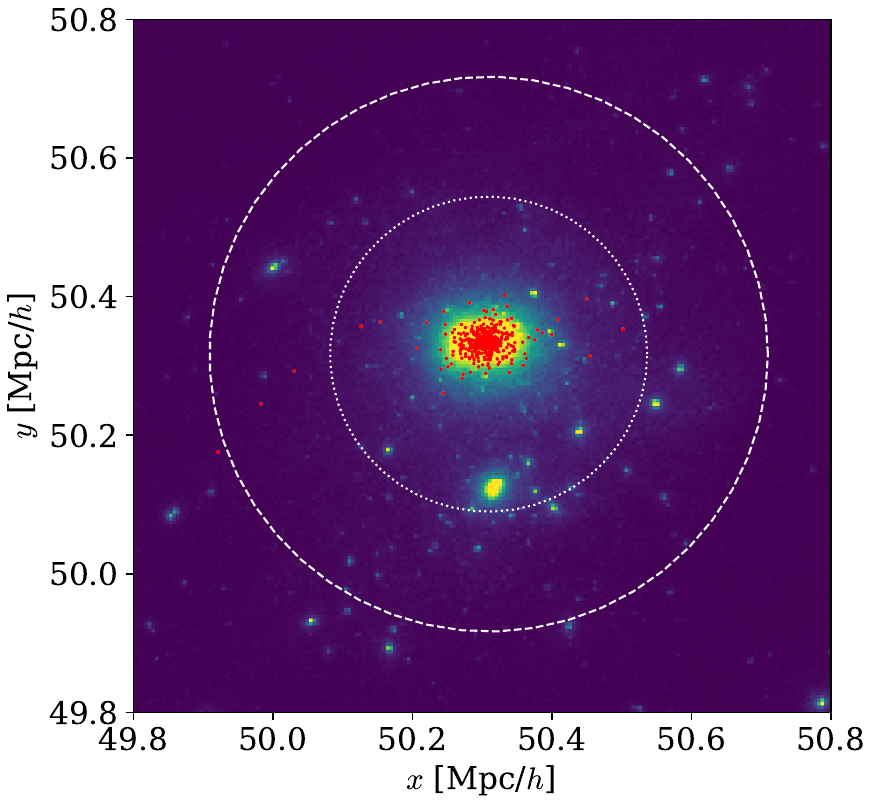}
    \caption{Cosmic string loop positions are overlayed on top of the DM density as described in \cref{fig:sim}. Each red point indicates the position of a loop, with only $\approx3$\% of loops displayed for readability. The left panel shows the case with zero rocket force, such that the loops are distributed like DM. The right panel shows the case for $\xi=12.5$, corresponding to the maximum of the right panel of \cref{fig:capture}, with a significant rocket force pushing most loops away from the halo. Also note that the smaller subhaloes cannot efficiently capture any such loops, while the larger main halo captures such loops successfully.}
    \label{fig:sim_rockets}
\end{figure}

The {\tt GADGET-4} code is run using appropriate initial conditions for a cosmological Zoom simulation which tracks the evolution of a Milky Way-sized galaxy, based on the Aquarius Project \cite{Springel:2008cc}. The simulation assumes a $\Lambda$CDM cosmology with $\Omega_m=0.25$, $\Omega_\Lambda=0.75$, and $H_0=73 \mbox{ km/s/Mpc}$. We simulate a volume with equal sides of comoving length $100$ Mpc$/h$, beginning at redshift $z_{\rm ini}=127$. The volume is populated with higher mass and lower mass dust grains such that the highest resolution is obtained for the target halo which evolves towards the center of the box by redshift $z=0$. The size of the halo can be characterized by $r_{x}$, which denotes the distance from the center of mass of the halo that encloses a density $x$ times greater than the critical (background) density. We shall use two such radii for later purposes, $r_{200}$ and $r_{50}$. For our target halo that resembles our own Milky Way halo, we have $r_{200} \approx 227$ kpc$/h$ and $r_{50} \approx 400$ kpc$/h$.\footnote{The assembly history and structural properties of the target halo are consistent with expectations for Milky-Way–like systems in $\Lambda$CDM cosmologies \cite{10.1093/mnras/staa2202}.
} For more information on this setup, see the {\tt GADGET-4} documentation (see \cref{gadget}). \Cref{fig:sim} shows the DM density in two dimensions (restricting to a $2 \mbox{ Mpc}$ range for the third dimension) for an example output of the simulation, along with the above two characteristic sizes highlighted.

On top of this cosmological DM simulation, we then add a population of the simulated string loop ``rockets". We perform many simulations where in each, we populate a fixed number of rockets (equal to $64^3$) for every $\xi$ uniformly over a large enough rectangular region (of comoving volume $336$ (Mpc/h)$^3$), that surrounds the final would be halo, and all the rockets have the same direction of the rocket force. We then repeat the same simulation over but with different values of the rocket force directions for every $\xi$.
We choose the following values of $\xi$ (i.e. rocket acceleration magnitude):
\begin{align}
    \xi = \{3, 4, 5, 6, 7, 8, 9, 10, 12.5, 15, 17.5, 20, 30, 40, 50, 60, 70, 80, 90, 100, 125, 150, 200, 300, 500, 1000\}\,.
\end{align}
For each value of $\xi$, we choose all six combinations of
\begin{align}
    \phi=\{0,\pi\}\qquad{\rm and}\qquad \cos \theta=\{-2/3,0,2/3\}\,,
\end{align}
for the rocket force direction, where $\phi$ and $\theta$ are azimuthal and polar angles respectively (with respect to one of the axis of the simulation). That is, for every value of $\xi$, we perform six simulations with $64^3 = 262,144$ rockets in each, and then finally combine their results. For later purposes, we therefore have
\begin{align}
    N_{\rm sim, tot} = 64^3 \times 6 = 1,572,864
\end{align}
rockets for every value of $\xi$ that are initialized over a comoving rectangular box of size $\approx 6.95$ Mpc/h uniformly. That is, the physical size of this initial population box is $6.95/(1+z_{\rm ini}) \approx 0.05$ Mpc/h, giving the simulated physical (uniform) density of initial loops to be 
\begin{align}
\label{eq:number_density_sim}
    n_{\rm sim, ini} = \frac{N_{\rm sim, tot}}{6 \times (0.05\,{\rm Mpc}/h)^3} \approx 1.6 \times 10^9\,({\rm Mpc}/h)^{-3}\,.
\end{align}
Comparing it to the actual physical (uniform) density of loops, \cref{eq:number_density_actual}, we see that for $G\mu = 10^{-12}$ and even $\xi = 1$, this is roughly an order magnitude larger. So we can be confident that we are simulating an adequate number of loops.\footnote{It could be argued that our angular distribution of rocket directions is quite coarse. While we anticipate that it should not affect our final aggregate results since we do not observe any special direction in the halo properties and surrounding regions, a more dedicated simulation set with lots of angular directions is needed for a definitive answer.}

The rockets are given no initial peculiar velocities in the simulation. In \cref{subsec:rocket} we contrasted the two components of a loop’s peculiar velocity—the rocket-induced acceleration and the initial velocity imprinted at formation. For instance, \cref{eq:xir} specifies, for a fixed \(G\mu\), the redshift ranges in which the rocket effect dominates over the initial-velocity contribution for loops of size \(\xi\). In our simulations, in order to explore a wide range of parameters, we will even include values of $\xi$ and $G\mu$ where this approximation breaks down; we have checked for the limiting cases (e.g. $G\mu=10^{-12}$ and $\xi=1000$) that ignoring the initial $v_0(t)$ of \cref{eq:v0_damped_damped} introduces only sub-percent errors in the final results of the number of loops captured and their final spatial distribution. Further, while at the initial redshift $z=127$ the rockets should have peculiar velocities accumulated from the rocket effect, c.f.~\cref{eq:vpec-rocket-dominated}, we ignore this initial velocity as well for ease in generating initial states for large numbers of simulations; again, we have checked that in the limiting cases (this time, $\xi < 10$) the error in final results is again only at the sub-percent level.

For comparison, and in addition to the above simulations, we ran one extra simulation with the rocket effect set to zero (the limit of infinite $\xi$). \Cref{fig:sim_rockets} shows the same DM density as \cref{fig:sim} with the positions of string loops overlayed in red. The left panel shows a simulation with no rocket force, such that the string loops are distributed like DM, while the right panel is for relatively large rocket force ($\xi=12.5$, corresponding to the maximum of the right panel of \cref{fig:capture}) with most of the loops thrown out of the DM halo. Comparing the two scenarios, one can also note that the smaller subhaloes are not able to efficiently capture any loops with $\xi=10$, while they are expected to capture loops of smaller rocket force. In the following section, we provide the results of these simulations and comment on their implications for the capture of cosmic string loops.

% ----------------- %
% ----------------- %
\section{Results}
\label{sec:results}
% ----------------- %
% ----------------- %

We now present the results of our simulations. Throughout this section, we quantify the extent to which cosmic string loops are captured and clustered in a Milky-Way–like halo by analyzing their positions at the final simulation output, $z = 0$. Rather than relying on energy-based diagnostics—which are subtle to interpret in a periodic simulation box\footnote{A detailed discussion of loop energies and their interpretation in a periodic volume is provided in Appendix~\ref{app:loop_energy}.}—we instead focus on two complementary and physically transparent diagnostics: a radius-based definition of capture and a comparison with a corresponding rocket-off baseline in which loops behave as dark-matter–like tracers without imparting gravitational back-reaction.

\subsection{Total number of loops captured in halos}
\label{sec:total_number_capture}

A natural working definition of capture in structure-formation studies is whether an object lies within the virialized region of a dark-matter halo at the final time. Motivated by this standard approach, and to enable a direct comparison with Ref.~\cite{Jain:2020dct}, we define a loop to be captured if it resides within a fiducial halo radius at $z = 0$. In Ref.~\cite{Jain:2020dct}, this criterion was used both to estimate the number of loops remaining inside the virialized region and to quantify the population added through secondary infall. Here we shall not try to distinguish between these two, and rather report on the total number of captured loops today that incorporates the full heirarchical structure formation history of our prototypical Milky-way-like halo.

In the astrophysical literature, halo boundaries are often characterized by radii $r_x$, defined such that the mean enclosed density equals $x$ times the present-day homogeneous matter density. Two commonly used choices are $r_{200}$ and $r_{50}$. Neither should be regarded as the unique or physically exact virial radius. The value $200$ originates as a rounded form of the top-hat spherical-collapse overdensity $18\pi^2 \simeq 178$, while $50$ is a lower overdensity threshold sometimes adopted in discussions of extended halo structure.\footnote{The $18\pi^2$ is obtained by taking the ratio of the top-hat density at the turnaround, and the background density at the time of collapse (when top-hat radius contracts to zero).} Both choices are conventional, and provide empirical characterizations of halo size.

Given the absence of a single, sharp boundary for a realistic halo—and in keeping with the methodology of Ref.~\cite{Jain:2020dct}—we examine loop capture using both $r_{200}$ and $r_{50}$. In our simulations, these two radii are similar in size relative to the region where loop clustering actually occurs, and they produce nearly identical capture fractions. For this reason, their primary role is to test the robustness of our conclusions to the precise choice of spatial cutoff, rather than to delineate distinct dynamical regimes. The broader cosmological picture of hierarchical structure formation still provides useful context—material outside the nominal virial radius can, in principle, be accreted through secondary infall—but in practice the loops of interest occupy a much more compact region. We therefore adopt this radius-based definition (using $r_{200}$ and $r_{50}$) as our operational criterion for determining whether loops are captured.

To interpret this radius-based capture definition, we compare each simulation with a corresponding ``rocket-off'' baseline in which the same loops evolve without the rocket force. In the rocket-off case, loops behave as passive dark-matter tracers: they respond to the gravitational potential of the halo but exert no back-reaction. This baseline allows us to determine how many loops would end up within the halo in the absence of recoil. The ratio of rocket-on to rocket-off captured loops therefore isolates the dynamical impact of the rocket force itself: values below unity indicate that recoil expels loops that would otherwise follow ordinary gravitational infall, while values closer to unity reflect trajectories increasingly similar to the dark-matter flow. With this capture definition established, we quantify how the rocket force affects loop retention as a function of the size parameter $\xi$. For each simulation, we identify the main halo at $z = 0$ and count the number of loops lying within the chosen radius, primarily $r_{200}$, and also $r_{50}$ for comparison. To isolate the impact of recoil, these counts are normalized by the corresponding number obtained in a rocket-off baseline simulation at the same $\xi$. In this way, the resulting capture fraction measures the proportion of loops that remain near the halo out of those that would have been retained in the absence of the rocket effect. Fractions significantly below unity indicate efficient rocket-driven ejection, whereas values near unity reflect behavior increasingly similar to dark-matter tracers.

\begin{figure}[h!]
    \centering
    \includegraphics[width=0.45\linewidth]{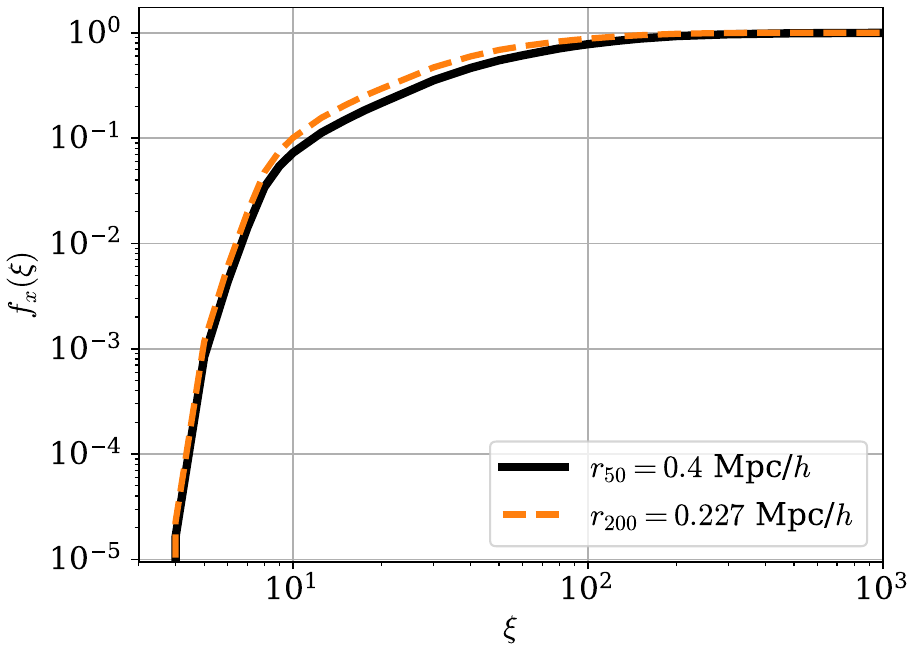}
    \includegraphics[width=0.45\linewidth]{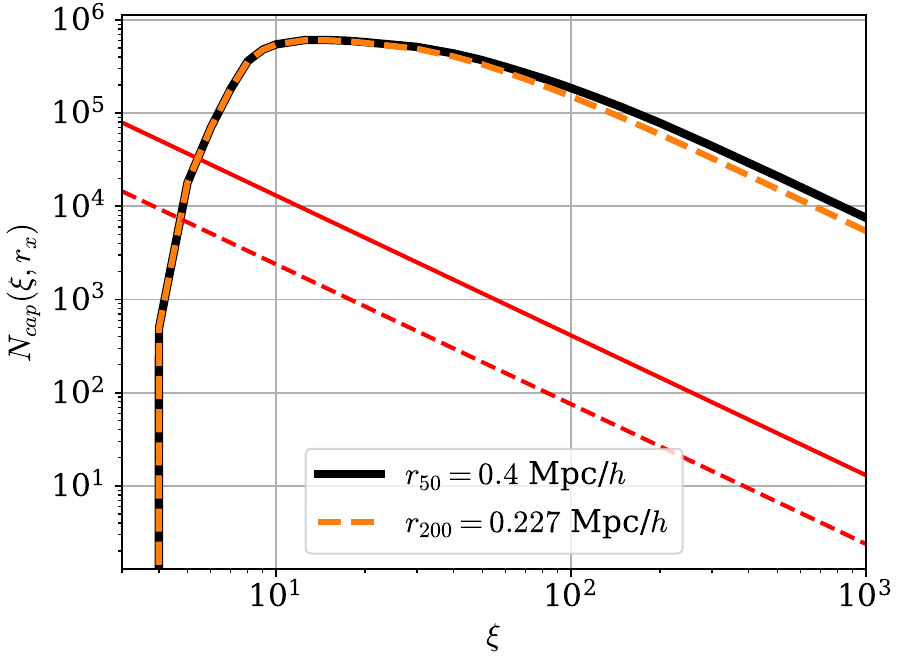}
    \caption{The fraction and number of loops captured within $r_{50} = 0.4 \mbox{ Mpc}/h$ (black solid curves) and $r_{200}=0.227~\mathrm{Mpc}/h$ (orange dashed curves) are shown. The left panel presents the relationship between $\xi$ and the fraction of loops captured by the galaxy. The fraction is a monotonically increasing function of $\xi$, rising more steeply for $\xi < 10$. The right panel shows the simulated count of captured loops, computed using the predicted number density in \cref{eq:number_density_actual} and assuming $G\mu=10^{-15}$; for different values of $G\mu$, this result can be rescaled by $(G\mu/10^{-15})^{-3/2}$. A peak is observed at $\xi \sim 15$. The red curves show the expected number of loops found within $r_{50}$ (solid red) and $r_{200}$ (dashed red), assuming a uniform distribution of loops and no gravitational capture.}
    \label{fig:capture}
\end{figure}

With $N_{\rm sim}(\xi,r<r_x)$ being the total number of $\xi$ sized loops within $r < r_x$ at $z=0$, while $N_{\rm sim}(f_{\rm rocket}\to0,r<r_x)$ being the total number of loops with zero rocket force (i.e. $\xi \rightarrow \infty$ formally) within $r < r_x$ at $z=0$, the resulting capture fraction $f_x(\xi)$ is
\begin{align}
    f_x(\xi) : = \frac{N_{\rm sim}(\xi,r<r_x)}{N_{\rm sim}(f_{\rm rocket}\to0,r<r_x)} \,,
\end{align}
with two different radii shown in the left panel of~\cref{fig:capture}. The overall trend is clear: for small values of $\xi$, where the rocket acceleration is strongest, only a small fraction of loops remain within the halo by $z = 0$. As $\xi$ increases and the rocket force becomes weaker, the fraction of retained loops rises steadily toward unity. This monotonic increase reflects that larger loops experience only mild recoil and therefore tend to follow trajectories increasingly similar to the rocket-off baseline. It is also clear that the two characteristic radii, $r_{200}$ and $r_{50}$ show no practical difference. That is, almost all of the captured loops are contained well within $r_{200}$. (This is also reflected in the right panel of~\cref{fig:sim_rockets}.)

The right panel of \cref{fig:capture} shows the predicted number of loops within the halo as a function of $\xi$, obtained using the analytic number density in \cref{eq:number_density_actual} and simulated number density in \cref{eq:number_density_sim}:
\begin{align}
    N_{\rm cap}(\xi,r_x) := \frac{n(\xi,z_{\rm ini})}{n_{\rm sim, ini}}\times N_{\rm sim}(\xi,r<r_x)\,.
\end{align}

Consistent with~\cite{Jain:2020dct}, we have the two anticipated features: a sharp rise in the total number of loops captured at small $\xi$, and a $\xi^{-3/2}$ fall off from the intrinsic loop abundance at large $\xi$. Their interplay produces a characteristic peak in the captured-loop population, which we notice to be around
\begin{align}
    \xi_{\rm peak} \simeq 12.5\,.
\end{align}
That is, the most abundant loops that are captured are of this size. For illustration, we also show the unperturbed -- due to Hubble flow if no structure formation had occurred -- number of loops with the two characteristic radii in red, which has the  $\xi^{-3/2}$ fall off. The presence of this $\xi_{\rm peak}$ is due to the fact that for smaller $\xi$ the rocket force imparts loops enough escape velocity for them to be able to escape the final resulting halo (and therefore expel them), while for larger $\xi$ the loops essentially behave as dark matter and the curve follows the otherwise unperturbed trend of $\sim \xi^{-3/2}$~\cite{Jain:2020dct}. 

In~\cite{Jain:2020dct}, a $\xi_{\rm min}$ was characterized which referred to the minimum value for which loops can be captured. For fiducial values of virialization radius and redshifts around $R_v = 60$ kpc and $z_v = 3$, we found it to be around 9 (see Eq.(34) there). Here we refrain from characterizing an $\xi_{\rm min}$ like that, since (a) it is rather difficult to extract good representative values for $R_v$ and $z_v$ in a true non-top-hat-like / hierarchical structure formation scenario, and (b) we do not simulate a variety of different haloes with different possible $R_v$ and $z_v$ values. A main feature of interest, consistent with~\cite{Jain:2020dct}, is the presence of this peak and an eventual $\xi^{-3/2}$ fall off.

In contrast, we note that the peak of $N_{\rm cap}$ is about $50-60$ times higher than what was estimated in~\cite{Jain:2020dct} (compare right panel of~\cref{fig:capture} with the blue curve of Fig.5 there). We anticipate that this apparent mismatch arises from the use of idealized spherical collapse and secondary infall models, which are known to be sensitive to virialization parameters and differ from the outcomes of hierarchical structure formation seen in simulations \cite{Gunn:1972sv,Bertschinger:1985pd,Suto:2016zqb}. Also notice that the number captured in terms of the virialization radius and redshifts (as estimated in Eq.(43) there), has a strong dependence on both -- it is proportional to $(1+z_v)^{15/2}$ and $R_v^{9/2}$. So a slight difference in $z_v$ (and/or $R_v$) can easily lead to orders of magnitude variations. So while it is not too surprising that we find an enhanced capturing of loops, it is nevertheless important and interesting from the point of view of observational signatures.

\subsection{Spatial distribution of captured loops}
\label{sec:spatial_distrib_capture}

\begin{figure}
    \centering
    \includegraphics[width=0.48\linewidth]{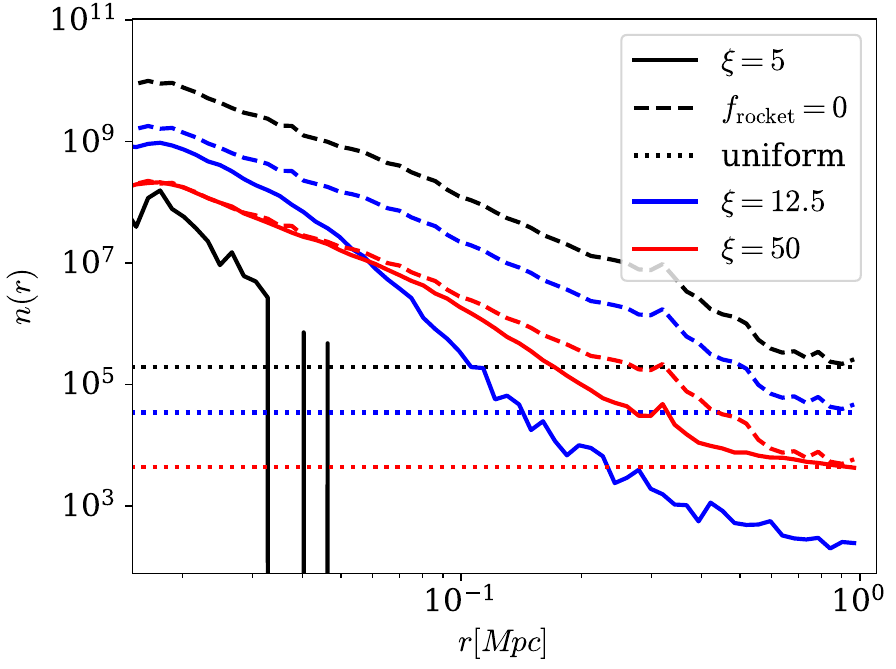}
    \caption{String loop number density is shown as a function of distance from the center of mass of the main halo. The dashed lines show the would-be number density for $\xi=15$ (black), $\xi=12.5$ (blue), and $\xi=50$ (red) in the absence of the rocket effect. The solid curves show the resulting distribution with the rocket effect present, indicating that the distribution tends to be more concentrated near the center of the halo for smaller $\xi$. Dotted lines lines of each color show the corresponding uniform number density for each $\xi$ in the absence of any gravitational clustering.
    }
    \label{fig:density_of_r}
\end{figure}

Beyond the overall count of cosmic string loops captured in a galaxy like ours, it is imperative to also study their spatial distribution. For instance, large loops without any significant rocket force (i.e. loops with with large enough $\xi$ values such that they can still be treated as point like particles), can be anticipated to cluster like dark matter particles. On the other hand, smaller loops for which the rocket force is significant but not too strong for them to be expelled out of the halo, can be anticipated to be clustered more towards the center as compared to outer regions of the halo, due to the stronger gravitational potential towards the center. For such loops, even if the rocket effect does not eject a significant proportion of them, all of the ones which are ejected are removed from the edges of the halo where the gravity is weakest, resulting in a centrally concentrated  distribution. 

This is precisely what we confirm in our simulations: larger loops tend to closely track the dark matter distribution, while smaller loops tend to be more concentrated towards the center of the halo. In \cref{fig:density_of_r}, we show the loop distribution $n_{\rm cap}(\xi, r_i)$ for three choices of length parameter $\xi$, defined as
\begin{align}
    n_{\rm cap}(\xi, r_i) := \frac{N_{\rm sim}(\xi, r_i - \Delta r < r < r_i + \Delta r)}{4\pi r_i^2\Delta r}\,,
\end{align}
where the distribution shown is discretized in bins $r_i$ of size $\Delta r$ which are logarithmically spaced in the range $[10^{-3},1] \mbox{ Mpc}/h$.
The figure shows the final distribution for $\xi=5$ (solid black), $\xi=12.5$ (solid blue), and $\xi=50$ (solid red). 
For a comparison with that of dark matter like behavior, we also show the corresponding dashed curves -- the would-be final distribution for the loops in the absence of the rocket effect ($f_\text{rocket} = 0$). The shape of these curves is exactly that of the dark matter density. Note that the three dashed curves differ in their amplitudes precisely because of the different number density for different $\xi$ -- the $\xi^{-3/2}$ dependence in~\cref{eq:number_density_actual}. Also shown with dotted lines are the background unperturbed number densities, the would-be density solely due to the Hubble dilution in the absence of any gravitational clustering. 

It is clear how the clustering perturbs this background density for different sized loops. In particular for the most abundant loops of size $\xi \sim \xi_{\rm peak} \approx 12.5$, their density towards the center is enhanced significantly compared to that of much larger loops (say of size $\xi \sim 50$) that track the DM density more closely, even though the $\xi_{\rm peak}$ loops are captured less efficiently. In contrast, the densities of the $\xi_{\rm peak}$ loops towards the outer regions of the halo are suppressed compared to the longer loops, as they are less tightly bound to the halo and thus the rocket effect efficiently ejects them. Concretely, from \cref{fig:density_of_r} we see that the density of $\xi_{\rm peak}$ sized loops, at the location of the Sun (which would correspond to the left edge of the plot) is $\sim 10^9$, which is about $\sim 10^4$ times higher than at $\sim 100$ kpc. This contrast is two orders of magnitude greater than the analogous ratio for loops of size $\xi \sim 50$.

% ----------------- %
% ----------------- %
\section{Conclusions}
\label{sec:summary}
% ----------------- %
% ----------------- %

The prospects for detecting cosmic strings through astrophysical and cosmological signatures depend sensitively on the spatial distribution of oscillating string loops in and around galaxies. A long-standing question is whether the rocket effect arising from anisotropic gravitational-wave emission prevents loops from being gravitationally captured—leading to an approximately uniform distribution—or whether a substantial population clusters within dark-matter haloes. In this work, we addressed this question using dark-matter–only zoom N-body simulations of the formation of a Milky-Way–like halo, supplemented with a population of effectively massless particles that model cosmic string loops subject to a constant rocket acceleration of magnitude $f_{\rm rocket}\propto (\xi t_0)^{-1}$ with fixed direction and no back-reaction on structure formation.

We quantified capture using a simple and robust operational criterion: a loop is deemed captured if it lies within a characteristic halo radius at $z=0$, focusing primarily on $r_{200}$ and verifying robustness with $r_{50}$. To isolate the dynamical impact of recoil, all capture counts were normalized to a corresponding rocket-off baseline in which loops behave as passive dark-matter tracers. The resulting capture fraction is a monotonic function of the loop size parameter $\xi$: strong rocket forces (small $\xi$) efficiently expel loops that would otherwise follow the halo flow, while for large $\xi$ the dynamics approach the rocket-off limit. The near-independence of the results on the choice of halo boundary indicates that essentially all captured loops reside well inside $r_{200}$.

Combining this capture efficiency with the intrinsic loop abundance $n(\xi)\propto\xi^{-3/2}$ yields a pronounced peak in the \emph{captured} loop population at $\xi_{\rm peak} \simeq 12.5$, reflecting the competition between rocket-driven ejection at small $\xi$ and the declining cosmological number density at large $\xi$. For fiducial string tensions $G\mu\simeq10^{-15}$, this corresponds to an absolute population of order $10^6$ loops bound within a Milky-Way–like halo. This number is larger than earlier analytic estimates based on spherical top-hat collapse \cite{Jain:2020dct}, highlighting the sensitivity of such estimates to idealized assumptions about halo formation and emphasizing the importance of hierarchical structure formation in determining the late-time loop population.

Beyond total counts, we find a qualitative dependence of the spatial distribution of captured loops on $\xi$. Loops with sufficiently weak rocket forces (large $\xi$) closely trace the dark-matter distribution. In contrast, loops subject to stronger recoil but still captured—particularly those near $\xi_{\rm peak}$ that dominate the bound population—are preferentially concentrated toward the central regions of the halo. For instance, their density near the Sun is about $10^4$ higher than in the inter-galactic region. This behavior arises because rocket-driven ejection operates most efficiently at large radii, while also reducing the loop density at all radii relative to dark matter, resulting in a surviving population that is centrally biased and significantly enhanced relative to a uniform (Hubble-flow) distribution.

These results have direct implications for observational prospects. Although loops with very small $\xi$ are efficiently expelled and those with very large $\xi$ are intrinsically rare, the dominant contribution to the local loop density arises from loops near $\xi_{\rm peak}$, which are both abundant and gravitationally bound. Since the Solar System resides well within the virial radius of the Milky Way, the enhanced central concentration of these loops implies that the local loop density can remain substantial despite partial depletion at large radii. As emphasized previously in \cite{Jain:2020dct}, this has important consequences for astrophysical and cosmological searches for cosmic strings—such as gravitational-wave emission from nearby loops~\cite{Xing:2020ecz,LISACosmologyWorkingGroup:2022kbp,CamargoNevesdaCunha:2022mvg,Suresh:2023hkz,LISAConsortiumWaveformWorkingGroup:2023arg} or other local signatures (such as due to supermassive black hole spin down~\cite{Ahmed:2024cty})—where estimates based on uniform loop distributions or purely analytic halo models may significantly underestimate the expected signal.

Several extensions would further refine the connection between loop microphysics and halo-scale observables. Our implementation models the rocket force as a constant acceleration with fixed direction and treats loops as non-shrinking point particles—choices that allow for a controlled assessment of recoil during structure formation but can be improved upon. Recent numerical studies of gravitational backreaction on cosmic string loops indicate that small-scale structure is smoothed by self-interaction, leading to refined gravitational-wave emission properties and quantitative corrections to standard radiation estimates \cite{Wachter:2024aos,Wachter:2024zly}. Incorporating these effects into the effective rocket-force modeling, along with exploring halo-to-halo variance using additional zoom simulations, would provide a more complete and observationally relevant characterization of cosmic string loop populations in galaxies.

\acknowledgments

We are especially grateful to Alex Vilenkin for originating the ideas that initiated this research program and for his foundational role in its early development. We also thank Neil Shah and Ken Olum for useful discussions. IJA is supported in part by NASA grant 80NSSC22K081 and by the Provost's Postdoctoral Society of Science Fellowship at the University of Notre Dame. During a significant portion of this work, MJ was supported by a Leverhulme Trust Research Project (Grant No. RPG-2022-145). Part of this work was conducted using computational resources and services at the Center for Computation and Visualization, Brown University.

\bibliography{bibfile}% Produces the bibliography via BibTeX.

\appendix

\section{Loop Energy Distributions at \texorpdfstring{$z=0$}{z=0}}
\label{app:loop_energy}

In this appendix we collect the energy-based diagnostics for string loops in our simulations. Although the total (kinetic plus gravitational potential) energy of each loop can be evaluated at any time, interpreting these energies in a periodic cosmological simulation requires care. In a periodic box, the gravitational potential does not asymptote to zero at large distances from the main halo. As a result, the sign of the total energy does not by itself provide an unambiguous criterion for determining whether a loop is gravitationally bound to the halo. For this reason, energy-based criteria are not used in the main analysis of loop capture in \cref{sec:results}; instead, energies serve primarily as a diagnostic for understanding loop dynamics in the rocket-off and rocket-on cases.

\begin{figure}[h!]
    \centering
    \includegraphics[width=0.5\linewidth]{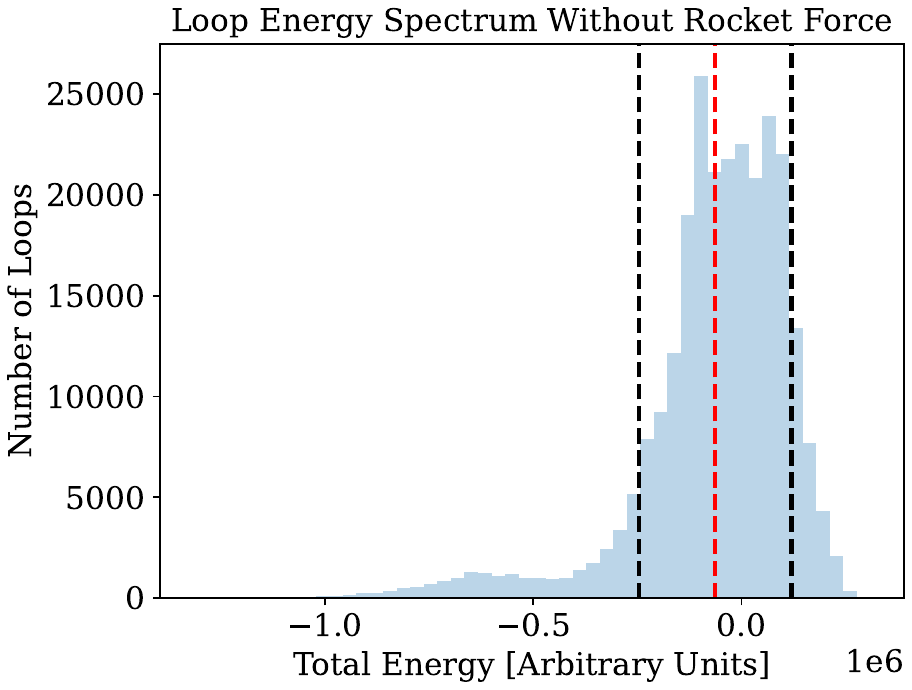}
    \caption{
    The energy spectrum of string loops in the final state at $z = 0$ is shown for the case with no rocket force. The mean loop energy (red dashed line) is below zero, with a distribution consisting mostly of negative energies but also including some positive-energy (black dashed lines indicate one standard deviation from the mean).}
    \label{fig:energy1}
\end{figure}

In the absence of the rocket force, loops behave as passive tracers of the dark-matter potential, experiencing only gravitational accelerations due to the evolving halo. Their final-state energies therefore provide a baseline against which rocket-on simulations may be compared. As seen in \cref{fig:energy1}, the distribution of loop energies in the rocket-off case peaks at negative values, with a substantial fraction of loops lying below zero. Positive-energy loops can also appear due to the periodic boundary conditions and the finite simulation volume.

\begin{figure}[h!]
    \centering
    \includegraphics[width=0.45\linewidth]{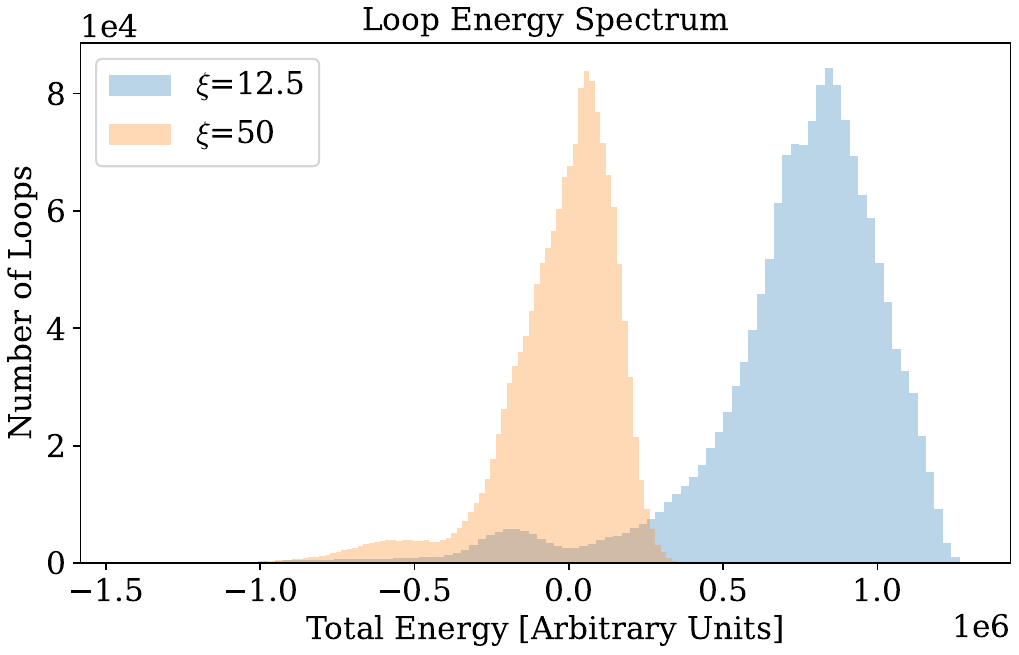}
    \includegraphics[width=0.45\linewidth]{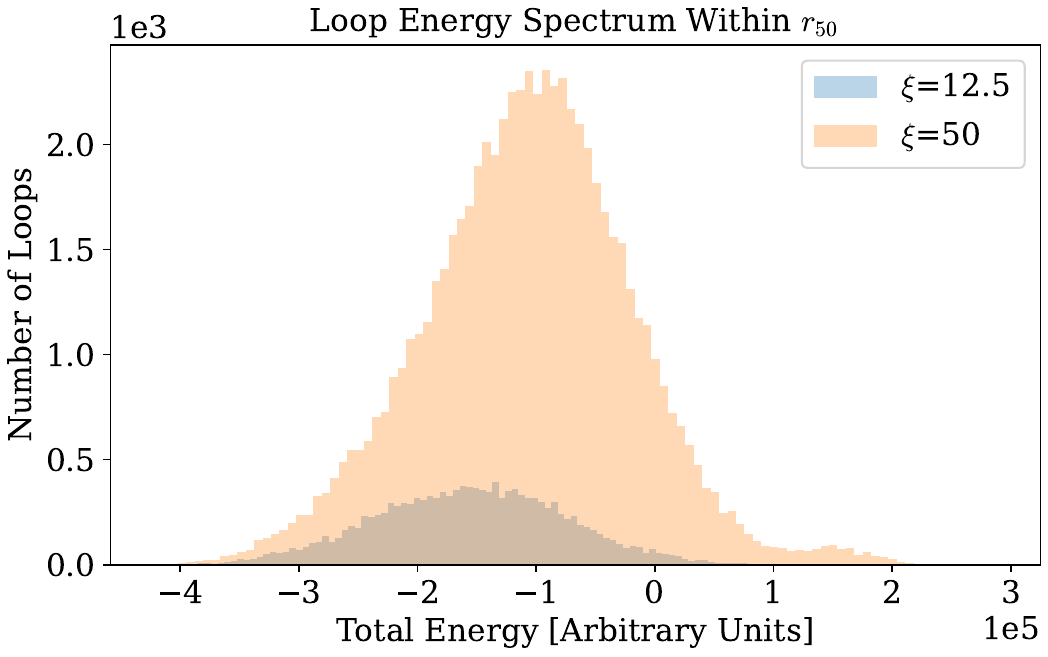}
    \caption{
    Final-state energy spectra at $z = 0$ for two values of the loop-size parameter, $\xi = 10$ (blue shaded region) and $\xi = 50$ (orange shaded region). The left panel shows all loops in the simulation box. Loops with smaller $\xi$ generally have higher kinetic energies due to the stronger rocket force, and therefore exhibit energies well above zero. Loops with larger $\xi$ are more tightly bound, with energies concentrated around negative values. The right panel shows only loops located within $r_{50}$ of the central halo. In both cases, loops within this region tend to have negative energies, indicating that those which remain near the halo are gravitationally bound. Loops with larger $\xi$ constitute a larger fraction of this population.}
    \label{fig:total_energy}
\end{figure}

\Cref{fig:total_energy} illustrates how the rocket force modifies the energy distribution. For small $\xi$, corresponding to large rocket acceleration, the kinetic energy dominates and the distribution shifts to high positive values. For larger $\xi$, where the rocket force is weak, the energy distribution resembles that of the rocket-off baseline and is concentrated near or below zero. Restricting attention to loops within $r_{50}$ of the halo center emphasizes that the loops which remain near the halo at $z = 0$ tend to occupy the negative-energy portion of phase space, consistent with their gravitational binding.

These energy spectra help illustrate how the rocket force affects loop dynamics, but they do not provide a robust or simulation-volume-independent definition of capture. For this reason, the main text relies instead on a radius-based capture criterion (loops within $r_{200}$ or $r_{50}$ at $z = 0$), calibrated against the corresponding rocket-off baseline. The energy-based results presented here serve as supplementary diagnostics and can assist in interpreting the behavior observed in \cref{sec:results}.

\end{document}